\documentclass[runningheads]{llncs}

\usepackage[dvipsnames]{xcolor}
\usepackage{cite}
\usepackage{todonotes}
\usepackage{amsmath}
\usepackage{amssymb}
\usepackage{ragged2e}
\usepackage[most]{tcolorbox}
\usepackage{hyperref}
\usepackage[capitalise,nameinlink]{cleveref}
\usepackage{tabularx}
\usepackage{multirow}
\usepackage{adjustbox}
\usepackage{colortbl}
\usepackage{threeparttable}
\usepackage{enumitem}
\usepackage{microtype}
\usepackage{bbding}
\usepackage{orcidlink}

\usepackage{siunitx}
\usepackage{tikz}
\usetikzlibrary{shapes.geometric, positioning, patterns}

\usepackage{booktabs}
\usepackage{xspace}

\usepackage{listings}
\usepackage[T1]{fontenc}  
\usepackage[scaled=0.75]{beramono}  
\usepackage{graphicx}
\usepackage[font=small]{caption}
\usepackage{longtable}
\usepackage{pdflscape}
\usepackage{subcaption}

\usepackage{array}
\newcolumntype{L}[1]{>{\raggedright\let\newline\\\arraybackslash\hspace{0pt}}m{#1}}
\newcolumntype{C}[1]{>{\centering\let\newline\\\arraybackslash\hspace{0pt}}m{#1}}
\newcolumntype{R}[1]{>{\raggedleft\let\newline\\\arraybackslash\hspace{0pt}}m{#1}}

\newcommand{\key}{KeY}
\newcommand{\smtlib}{SMT-LIB}
\newcommand{\cvc}{cvc5}
\newcommand{\zthree}{Z3}
\newcommand{\bitwuzla}{Bitwuzla}
\newcommand{\vc}{VC}

\definecolor{cbred}{HTML}{D55E00}
\definecolor{cborange}{HTML}{E69F00}
\definecolor{cbyellow}{HTML}{F5C710}
\definecolor{cbgreen}{HTML}{009E73}
\definecolor{cbcyan}{HTML}{56B4E9}
\definecolor{cbblue}{HTML}{0072B2}
\definecolor{cbpurple}{HTML}{CC79A7}

\newcommand{\eg}{\emph{e.g.,}}
\newcommand{\ie}{\emph{i.e.,}}

\newcommand{\code}[1]{\lstinline{#1}}

\newcommand{\TightenPar}{\looseness=-1}

\newcommand{\circled}[1]{\tikz[baseline=(char.base)]{\node[shape=circle,fill={rgb,255: red,217; green,217; blue,217},inner sep=1pt] (char) {#1};}}

\newtcolorbox{rqanswer}[1][]{
    colback=gray!10!white,
    colframe=gray!50!black,
    boxrule=1pt,
    arc=2mm,
    left=4pt,
    right=4pt,
    top=0pt,
    bottom=0pt
}

\newif\ifarxiv
\arxivtrue

\makeatletter
\newcommand{\printfnsymbol}[1]{%
  \textsuperscript{\@fnsymbol{#1}}%
}
\makeatother

\begin{document}

\title{Verifying Floating-Point Programs in Stainless}
\date{}

\author{Andrea Gilot\thanks{These authors contributed equally to this work.}\Envelope{} \orcidlink{0009-0006-4463-9414} \and
Axel Bergstr\"{o}m\printfnsymbol{1} \orcidlink{0009-0009-2968-9167} \and
Eva Darulova \orcidlink{0000-0002-6848-3163}}
\institute{Uppsala University \\
\email{\{andrea.gilot,axel.bergstrom,eva.darulova\}@it.uu.se}}

\authorrunning{A. Gilot, A. Bergstr\"{o}m and E. Darulova}

\maketitle

\begin{abstract}
    We extend the Stainless deductive verifier with floating-point support,
    providing the first automated verification support for floating-point
    numbers for a subset of Scala that
    includes polymorphism, recursion and higher-order functions.
    We follow the recent approach in the KeY verifier to axiomatise reasoning
    about mathematical functions, but go further by supporting all functions
    from Scala's math API, and by verifying the correctness of the axioms
    against the actual implementation in Stainless itself.
    We validate Stainless' floating-point support on a new set of benchmarks
    sampled from real-world code from GitHub, showing that it can verify
    specifications about, \eg{} ranges of output or absence of special values
    for most supported functions, or produce counter-examples when the specifications
    do not hold.
\end{abstract}

\section{Introduction}
\label{sec:introduction}
Floating-point arithmetic is ubiquitous in modern software. 
While it is often discussed in the context of scientific computing, embedded systems, and machine learning, it is first and foremost pervasive in everyday user code~\cite{gilot2025largescalestudyfloatingpointusage}.
Yet, 
verification tools do not always support floating-point reasoning, limiting their use in practice.
For example, the absence of floating-point support in Liquid Haskell~\cite{vazou2014refinement} has been identified as a limitation to its applicability in the real world~\cite{vazou2014liquidhaskell}.
\TightenPar{}

This work introduces bit-accurate floating-point reasoning into Stainless~\cite{hamzaSystemFR2019}, a deductive verification framework for Scala programs.
Two main challenges in this integration are correctness and performance. 
To ensure correctness, the implementation must carefully address semantic discrepancies between the JVM and SMT-LIB,
and the non-standard behaviour of equality.

Moreover, reasoning about floating-point arithmetic in SMT solvers is known to be expensive, requiring a careful encoding of verification conditions
using specialised \smtlib{} constructs when available.

We start by reproducing the floating-point support recently introduced in \key{}~\cite{abbasiCombiningRuleSMTbased2023}, demonstrating that this approach can be generalised to other verifiers.
Our integration extends this support to the subset of the Scala language handled
by Stainless, enabling the verification of programs that combine floating-point
computations with, among others, higher-order functions and algebraic data types, 
features unsupported in \key{}.
We address the challenge of combining polymorphic equality, often assumed to be reflexive, with floating-point reasoning, where this assumption does not hold (because \code{NaN != NaN}).
Ignoring this mismatch can lead to unsoundness, as theorems proven for polymorphic types may be unsafely instantiated with floating-point types.

We implement a standard math library for Stainless that mirrors the Scala
standard library API, and verify specifications \emph{in Stainless itself} that
were previously only axiomatised in \key{} or not supported at all.
While these specifications are partial, \eg{} encoding that \lstinline{sin}
returns a value between -1 and 1, they are sometimes sufficient for verifying user
code with transcendental functions. SMT-solvers do not support transcendental
functions, and inlining the implementation of math library functions will
typically yield unacceptably long solving times.



To minimise user annotation effort, Stainless \emph{automatically} generates
verification conditions that capture subtle aspects of floating-point behaviour
that are counter-intuitive and often misunderstood, even among experienced
developers~\cite{dindaDoDevelopersUnderstand2018}.
Specifically, it enforces non-\code{NaN} preconditions for floating-point
comparisons and verifies the safety of casts between floating-point and integer
types.
\TightenPar{}

We evaluate Stainless on the benchmarks proposed by \key{}~\cite{abbasiCombiningRuleSMTbased2023}, showing comparable support for floating-point reasoning.
We further assess Stainless on 103 real-world floating-point functions mined from open-source GitHub repositories, identifying both practical use cases and remaining challenges---for Stainless and beyond---for achieving comprehensive floating-point verification support.

Our results show that Stainless successfully verifies all benchmarks from \key{} and either verifies or produces counterexamples for 82\% of user benchmarks.
Reasoning about transcendental functions via properties verified in our mathematical library is sufficient for more than half of these user benchmarks, highlighting both its practical effectiveness but also its current limitations.
Finally, automated checks for unintuitive floating-point behaviour detect every occurrence of the issues checked for in user code, effectively identifying potential bugs.

\paragraph{Contributions.} In summary, we make the following contributions:

\begin{itemize}
    \item We extend the Stainless verifier with floating-point support\footnote{Merged into the main Stainless repository (\url{https://github.com/epfl-lara/stainless}).}, the recent SMT-solver Bitwuzla~\cite{niemetzBitwuzla2023} as a backend specialising in floating-point reasoning, and usability features such as automatic checks for unintuitive floating-point behaviour (\Cref{sec:implementation}).
    \item  We implement a standard math library for Stainless that closely follows the OpenJDK math library implementation and verify mathematical properties of this library with Stainless itself (\Cref{sec:case_study}).
    \item We evaluate Stainless on the set of benchmarks proposed by \key{} as well as on user-written floating-point code mined on GitHub\footnote{\url{https://github.com/fxpl/stainless-float-benchmarks}} (\Cref{sec:evaluation}).
\end{itemize}


\section{Overview}
\label{sec:overview}



\begin{figure}[t]
\begin{lstlisting}
case class StormDay(moves: Int, errors: Int) {
  require(moves >= errors && errors >= 0 && moves > 0)  // class invariant

  def accuracyPercent: Float = {
    100 * (moves - errors) / moves.toFloat
  }.ensuring(result => 0 <= result && result <= 100)  // postcondition
}

def gradient(prediction: Double, label: Double): Double = {
  require(prediction.isFinite && label.isFinite)  // precondition
    - 4.0 * label / (1.0 + math.exp(2.0 * label * prediction))
}.ensuring(result => !result.isNaN)

def limit(maxMagnitude: Double): Vector = {
  if (getMagnitude > maxMagnitude) // NaN check inserted here
    getNewMagnitude(maxMagnitude)
  else Vector(x, y)
}
\end{lstlisting}
\caption{Example floating-point programs extracted from GitHub} \label{lst:examples}
\end{figure}

Stainless did not previously support floating-point reasoning.
In this section, we provide an overview of Stainless and of the floating-point support introduced in this work.
Stainless is a contract-based verifier for Scala programs \cite{hamzaSystemFR2019}. Every Stainless program is valid Scala code and can be compiled and executed on the JVM like any other Scala program.
The class \code{StormDay} in \Cref{lst:examples} illustrates a fragment of the Lichess codebase, an online chess platform.
The function \code{accuracyPercent} computes a player's accuracy by counting the number of correct moves in a sequence of chess moves.
Under the assumption that the number of moves is strictly positive and greater
than the number of errors (expressed in the class invariant in the
\code{require} clause), we can specify in the postcondition
(\code{ensuring} clause) that the result must lie within the interval $[0, 100]$.

During verification, Stainless generates verification conditions that are discharged by SMT solvers.
In this case, Stainless reports the counterexample \code{StormDay(1073741832, 730144766)}, for which the function returns \code{-2.9586256E-5}, violating the postcondition.
This result occurs because the left multiplication uses integer arithmetic, overflows to a negative value, and is then divided.
Rewriting the body of \code{accuracyPercent} 
to \code{100 * ((moves - errors) / moves.toFloat)} forces the floating-point division to occur before the multiplication by 100, guaranteeing that the result always lies within the expected range.


The function \code{gradient} in \Cref{lst:examples} implements a gradient computation in Apache
Spark~\cite{apacheSpark}, a Scala framework for large-scale data processing and
the most popular Scala library on GitHub.
For this example, we specify that the inputs are finite, \ie{} they are
neither of the floating-point special values \code{+/-Infinity} nor \code{NaN}. One
can obtain infinite floating-point values due to, \eg{} overflow and ``Not a
Number'' values due to mathematically invalid operations, \eg{} \code{sqrt(-1.0)}.
Stainless attempts to verify that the function never returns \code{NaN}, which one
can view as saying that it does not encounter an error.
When run on this snippet, Stainless reports a counterexample in which
\code{label} is assigned a value of very large magnitude (causing overflow when
multiplied by $2$) and \code{prediction} is set to $0$, leading to \code{Infinity * 0 = NaN}.
Again reassociating the multiplications to the right solves the problem and ensures compliance with the specification, as verified by Stainless.

This example poses two challenges for the verifier: the detection of \code{NaN} values and the presence of the exponential function.
Detecting \code{NaN} values is difficult for both developers and static analysers because their precise characterisation requires bit-level reasoning.
The difficulty increases in the presence of transcendental functions,
\code{math.exp} in this example, that are not natively supported by SMT solvers
and whose implementations are typically too complex to be inlined.
\TightenPar{}


Stainless supports reasoning about mathematical functions from the Scala standard library such as trigonometric, exponential or logarithmic functions.
Following the approach of \key{}~\cite{abbasiCombiningRuleSMTbased2023}, these functions are modelled as axiomatised uninterpreted symbols.
This abstraction enables Stainless to perform basic reasoning about the occurrence of special values such as \code{NaN}, without requiring complete semantic modelling. Unlike KeY, we verify the axioms against the actual implementation with Stainless itself---stress-testing its floating-point support as these functions are highly numerical in nature; \Cref{sec:case_study} provides more details.


Special-value bugs frequently arise from missing \code{NaN} checks \cite{diFrancoAComprehensiveStudy2017}.
For example, NumPy previously contained a bug in its \code{max} function, which returned an arbitrary number instead of the maximum when \code{NaN} values were present in the input array.
The cause for such bugs is \code{NaN}'s unintuitive behaviour: any comparison involving \code{NaN} evaluates to false, including \code{NaN == NaN}.
To prevent such errors from going undetected, Stainless automatically inserts \code{NaN} checks for all comparisons and equality operations, ensuring that neither operand is \code{NaN}.

The function \code{limit} in \Cref{lst:examples} clamps the magnitude of a 2D vector if it exceeds a given threshold.
If the input threshold is \code{NaN}, the comparison evaluates to false, causing the function to return the original vector, effectively ignoring the invalid input.
With the inserted \code{NaN} checks, Stainless automatically detects such cases and reports a corresponding counterexample.
When comparisons with \code{NaN} are intentional, users can explicitly disable these checks.

Stainless also inserts checks when casting floating-point values to integers that verify that input arguments are in the correct range.
In Scala, casting \code{NaN} yields $0$, while out-of-range values are clamped to the maximum or minimum representable 32-bit or 64-bit integer.
For 16-bit and 8-bit integers, however, out-of-range values wrap around modulo the target range, introducing a further source of unintended behaviour.

Most of the properties checked for in our examples and evaluation benchmarks involve range analysis, sign determination, and detection of special values. Floating-point equalities, which Stainless can also reason about, are less common.
This is mainly because expressions that are equal over real numbers (in an ideal specification), will in general not be equal over floating-points due to rounding.

In addition to these high-level floating-point properties, Stainless can also reason about bit-level properties of floating-point values.
Our case study in~\Cref{sec:case_study} requires reasoning about, for example, how bit-level operations on the exponents affect the corresponding floating-point values.
\section{Implementation of Floating-Point Support in Stainless}
\label{sec:implementation}

\begin{figure*}[t]
    \centering
    \includegraphics[width=0.95\textwidth]{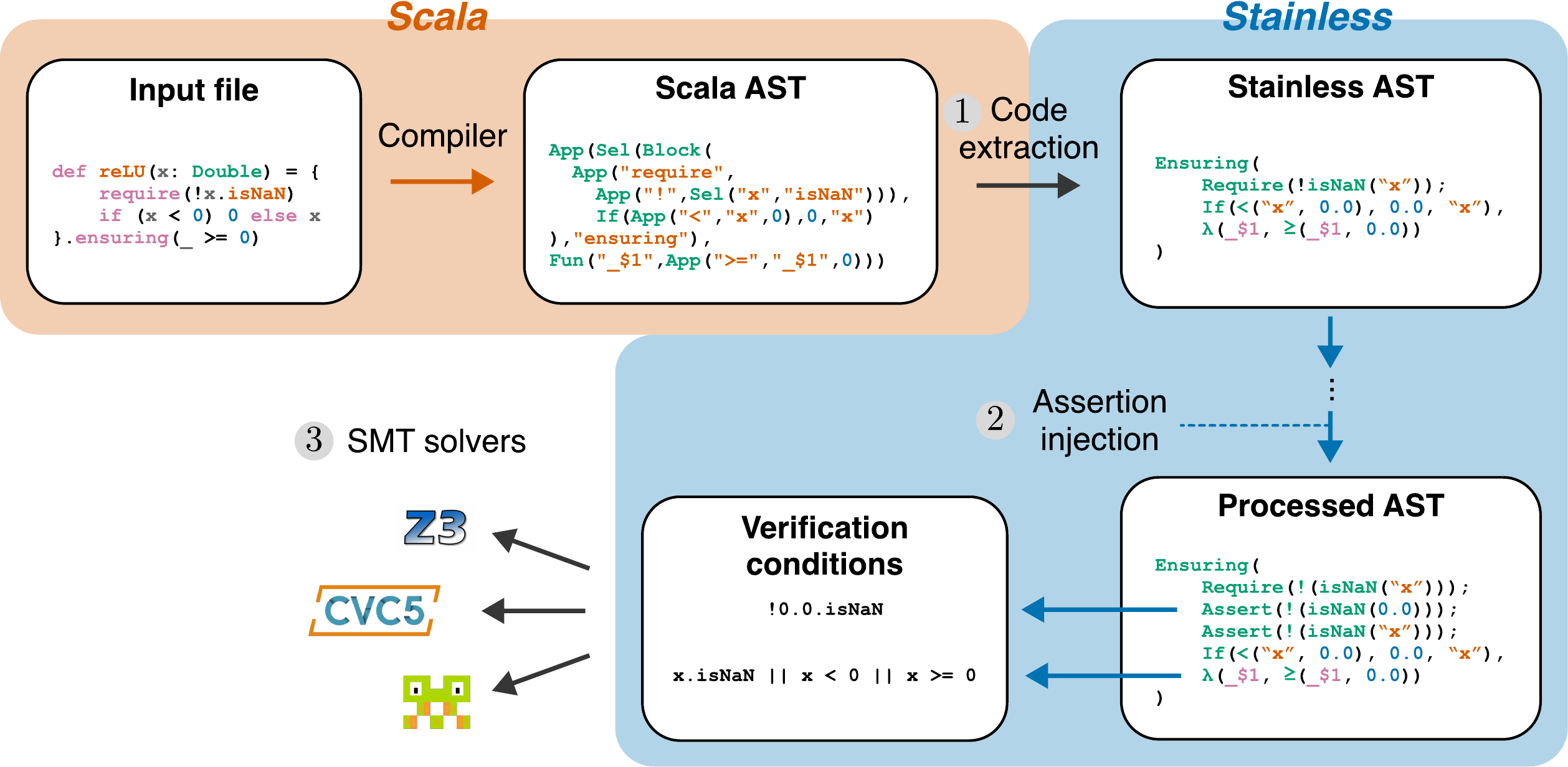}
    \caption{Simplified architecture of Stainless
    }
    \label{fig:architecture}
\end{figure*}

We implement our extension within Stainless, which is implemented in Scala, and reuse existing infrastructure wherever possible.
\Cref{fig:architecture} presents an overview of Stainless.
Stainless operates on the ASTs produced by the Scala compiler. It then processes the AST in several phases including anti-aliasing, type encoding and termination measure inference. 
At the end of the pipeline, Stainless generates verification conditions and discharges them to SMT solvers.
Note that the Scala compiler preserves floating points semantics (modulo compiler implementation bugs), and does not affect the verification results due to changes in floating-point rounding.
Our extension reuses most of the existing infrastructure and brings modifications to only a few phases in the pipeline, thereby showing Stainless infrastructure's modularity.

We now expand on the modifications we made to Stainless to support
floating-point arithmetic both soundly and efficiently that we deem to be the most interesting for other
tools and developers.

\paragraph{\textbf{Code Extraction.}}(\circled{1})
Part of the process consists in lowering Scala ASTs into Stainless ASTs.
Scala provides high-level constructs, while Stainless uses an intermediate representation tailored for verification, which requires explicit encodings.
By default, Stainless inlines non-recursive expressions and represents primitive operations such as arithmetic, comparisons, and type casts with dedicated nodes.
However, \smtlib{} defines more complex operators such as square root, min/max, and rounding (\code{ceil}, \code{floor}, \code{rint}). 
Mapping Scala calls directly to these operators reduces the size and complexity of verification conditions, so
we disable inlining for such functions and translate them directly.
For example, Scala implements the floating-point absolute value function by converting the input into binary, applying a mask, and reconstructing a floating-point value. 
Instead of inlining this bit-level implementation, we map calls to \code{abs} directly to the corresponding \smtlib{} operator.

Semantic discrepancies between the JVM and \smtlib{} also pose challenges.
One such discrepancy concerns the modulo operator: in Scala \code{14.5 \% 1.5} evaluates to \code{1.0}, while in \smtlib{} it yields \code{-0.5}.
Correcting this difference (e.g., by adding an offset) introduces rounding errors, so Stainless does not currently support floating-point modulo. 
To assess the impact of this limitation, we analysed 30,421 floating-point functions mined from GitHub (\Cref{sec:emp_eval}).
Only 108 (0.35 \%) used modulo, typically with arguments 1 (to extract the fractional part) or 360 (for angle computations).
In these cases, if one is willing to accept different floating-point rounding behaviour, one can replace \code{x \% n} with \code{x - n * floor(x / n)} to avoid the modulo operation
(although both expressions are mathematically equivalent over the reals, they accumulate rounding errors differently in floating-point arithmetic).
Since Scala runs on the JVM, \key{} is subject to the same limitation and likewise omits support for floating-point modulo.

A second discrepancy concerns the representation of \code{NaN} values.
In IEEE 754, \code{NaN}s are encoded with all exponent bits set and one or more non-zero significand.
This yields multiple encodings of \code{NaN} values where the varying significand bits form a payload.
The JVM preserves these payloads, while \smtlib{} uses a single canonical \code{NaN} value. 
As a result, functions that expose the binary encoding of floating-point numbers diverge in semantics,
since \smtlib{} does not differentiate between different binary \code{NaN} encodings.
To avoid unsoundness, Stainless models binary conversion as an uninterpreted function whose result,
when converted back to a floating-point value, equals the original floating-point value.

\paragraph{\textbf{Equality and Polymorphism.}}
The \smtlib{} standard distinguishes between bitwise equality and IEEE754 equality, where \code{+0} is equal to \code{-0} and \code{NaN} is not equal to anything, including itself.
The non-reflexivity of equality is a problem when mixed with parametric polymorphism.
In Stainless, reasoning about type variables relies on the assumption that equality is reflexive, symmetric, and transitive.  
When such type variables are instantiated with floating-point types or with user-defined types that override equality, these assumptions may lead to unsoundness.  
For instance, Stainless could incorrectly prove \code{NaN == NaN}.

One possible solution is to drop all assumptions about parametric equality and instead treat it as an uninterpreted symbol, as done in Flux~\cite{lehmannFluxLiquidTypes2023}.  
Although this guarantees soundness, it breaks backward compatibility because many programs that previously verified in Stainless would no longer do so.  

We adopt a more pragmatic approach.  
By default, Stainless only instantiates type variables with types whose equality satisfies reflexivity, symmetry, and transitivity.  
If the user wishes to instantiate a polymorphic expression with \texttt{Float} or \texttt{Double}, they must annotate the type variable with \code{@noeq}.
Stainless then treats equality on the annotated type variable as an uninterpreted predicate, thereby avoiding unsoundness due to assumed algebraic properties of equality.

Even disregarding \code{NaN}s, IEEE754 equality does not necessarily imply behavioural equivalence either.
For example, \code{+0.0} and \code{-0.0} compare as equal, even though \code{.isPositive} returns true for the former and false for the latter.
Stainless handles this behaviour correctly.

\paragraph{\textbf{Assertion Injection.}}(\circled{2})
One of the phases we most extensively modified in our work is assertion injection.  
For integer arithmetic, Stainless inserts checks by default to ensure that operations do not overflow and that divisions by zero do not occur.  
This behaviour can be disabled globally through command-line options or locally via wrappers. Assertions are injected during AST traversal, depending on the node being processed.

We extend this mechanism with additional checks for floating-point operations.  
In particular, we add \code{NaN}-checks on comparisons and safety checks for casts to integers.
When processing a comparison node (including equality), Stainless now generates a verification condition ensuring that both operands are non-\code{NaN}.
When traversing a type cast to an integer, Stainless generates two verification conditions: one ensuring that the value is not \code{NaN}, and another ensuring that the floating-point value lies within the integer domain.
As demonstrated in our empirical evaluation (\Cref{sec:eval-rq3}), these checks can detect real bugs in user code.

It could be argued that NaN values should be checked globally, \eg{} by disallowing them in any function argument.  
While appealing from a correctness perspective, such control would raise concerns about the usability of the tool. 
Excessive checks can overwhelm the SMT solver and substantially increase verification time.
It is also unclear what assumptions these checks should make about the inputs of checked functions or expressions.
In addition, \code{NaN} values may be acceptable in some contexts (e.g., to describe out-of-domain behaviour).
In such cases, users should retain the ability to disable these checks on a per-function basis.
Another approach would be to provide fine-grained control, for example by supporting different levels of strictness or configurable options.  
Evaluating such usability trade-offs requires a user study, which we leave for future work.

\paragraph{\textbf{Addition of \bitwuzla.}}(\circled{3})
We extend the set of available SMT solvers in Stainless to include \bitwuzla{}~\cite{niemetzBitwuzla2023}.
\bitwuzla{} offers strong native support for floating-point arithmetic and bit-vectors, and outperforms other solvers on these types of queries (\Cref{sec:eval-rq2}).
However, it lacks support for algebraic data types (ADTs), which are heavily used in Scala programs.
For verification conditions involving ADTs, Stainless falls back to general-purpose solvers such as \zthree{}~\cite{deMouraZ32008} or \cvc{}~\cite{barbosaCVC52022}.
\TightenPar{}


\section{Case Study: Verified Standard Math Library}
\label{sec:case_study}

Floating-point programs often rely on standard mathematical libraries, particularly for transcendental functions \cite{gilot2025largescalestudyfloatingpointusage}.
However, implementations of these functions are currently often too complex for SMT solvers to reason about directly.
Instead, the \key{} verifier for Java axiomatises them~\cite{abbasiCombiningRuleSMTbased2023}, exposing only selected properties such as their output range or behaviour on special input values.
We adopt this property-based approach for Stainless, avoiding inlining low-level implementations and keeping verification tractable for programs that depend on them.
\TightenPar{}

Stainless goes further than \key{} by verifying that these exposed properties hold for implementations equivalent or close to the OpenJDK math library,
which is based on the widely used FdLibm C library~\cite{fdlibm}. 
Specifically, we implement a verified mathematical library in Stainless that mirrors the Scala standard library API.
Non-transcendental functions use the existing Scala implementation and are inlined during verification. 
We extend the set of supported transcendental functions beyond those axiomatised in \key{}, covering the full Scala math library, including hyperbolic functions, logarithms, cube roots, and the hypotenuse.
We also verify additional properties such as sign preservation and relations to the identity function.
\ifarxiv
\Cref{tab:key_vs_stainless_axioms} in the appendix
\fi
 summarises and compares the properties supported in \key{} and Stainless.
Users can choose whether Stainless should reason about transcendental functions through these verified properties or whether it should inline their implementations. 

For most transcendental functions (15 out of 18), the implementation in Stainless adapts FdLibm's code into Single Static Assignment (SSA) form, in which each variable is assigned exactly once.
This conversion eliminates mutability and allows us to bypass Stainless's encoding of mutable state, which would otherwise inflate verification conditions and degrade solver performance.
We perform the SSA transformation manually and validate the resulting code by testing it against the original implementation on ten million random inputs, stratified by magnitude, as well as on special values such as zeros, infinities, \code{NaN}s, and multiples of~$\pi$.
We also oversample subnormal numbers, as they have been shown to be a common source of bugs in floating-point code~\cite{diFrancoAComprehensiveStudy2017}.
As Stainless contracts are compiled by default to runtime assertions, fuzzing the implementations also checks if the proven properties hold for the tested inputs.

\subsubsection{Trigonometric Functions.}

The trigonometric functions \code{sin}, \code{cos}, and \code{tan} are implemented as two distinct steps:
1. a range reduction computing the input modulo $\pi/2$, and
2. the computation of the function value on the reduced argument.
Properties of the second step can be verified using the same approach as for the other FdLibm functions.
Some more extensive modifications were required to show that the first step returns a value in the expected range, due to poor SMT-solver performance.

The remainder modulo $\pi/2$ (RemPio2) operation has cases for infinite or \code{NaN} values, small inputs, intermediate inputs (magnitude $\sim\leq 2^{20}\pi/2$), and a final case that can handle all finite floating-point numbers.
Stainless can easily handle the first two cases, but we did not manage to complete the proofs for the intermediate magnitude case due to poor SMT-solver performance.
Instead, we remove the intermediate range case from our implementation, using the algorithm from the final case also for intermediate inputs.
We describe the algorithm for this final case and some of the modification we made for the proof.

It is not possible to compute the remainder using ordinary floating-point operations, since this would yield unacceptably low accuracy.
Instead, the algorithm represents values using arrays of either floating-point values or integers to achieve greater precision than ordinary double-precision floats.
After converting the input to this form, it is:
1. multiplied by a (input-magnitude-dependent) part of the expansion of $2/\pi$;
2. has the integer part removed;
3. is multiplied by a prefix of the expansion of $\pi/2$; and,
4. is compressed from an array to a double-precision float with a compensation term (another double-precision float containing the part of the result too small in magnitude to be represented by the first one).
After our modifications, Stainless can show that this result is in the expected range of $[-\pi/4, \pi/4]$ in about 20 minutes.

Arithmetic operations on these arrays are implemented using loops.
Since Stainless is a functional verifier, it does not support loops natively.
Instead, Stainless converts loops to recursive functions that return a tuple containing the updated values of all variables modified by the loop.
In SMT-LIB, these tuples are encoded as algebraic data types.

This means we cannot rely on the high performance of \bitwuzla{}, which does not support algebraic data types.
Instead, we have to help \cvc{} solve the verification conditions by adding extra assertions and sometimes unrolling a few loop iterations.
Both of these are done to improve solver performance.
We also modify the conversion of the input to the array form, ensuring that the same array index always represents values of the same magnitude.
This simplifies the loop invariants from a disjunction of different cases to a single case.

The original implementation uses an adaptive number of bits of the expansion of $2/\pi$,
re-doing part of the computation with more bits if the precision is found to be insufficient.
Since it would be very challenging, in Stainless, to show that this process terminates,
we simplify the implementation by using a constant number of bits of $2/\pi$.

Finally, we also modified the loop implementing the compression in step 4 to admit a simple invariant.
Our implementation of the loop is based on Algorithm~4.1 in Ogita et. al.~\cite{OgitaTakeshi2005}, but uses the \code{FastTwoSum} function instead of the \code{TwoSum} function.
Together, these modifications allow us to show that the result is in the expected range,
and therefore show that the trigonometric functions yield values in the expected ranges.


We tested the trigonometric functions on \emph{all} single-precision floats and 10 billion random double-precision values, in addition to the fuzzing described earlier.
Since the implementation is no longer bit-by-bit equivalent to the original, we have to modify the way we test the trigonometric functions.
Stainless' implementation of trigonometric functions is within 1 unit in the last place (ulp) of the OpenJDK implementation on all tested inputs.


\section{Evaluation}
\label{sec:evaluation}

We evaluate our extension of Stainless on three sets of benchmarks to investigate the following research questions:

\textbf{RQ1:} How effectively does Stainless verify the correctness of real-world floating-point programs or generate counter-examples when verification fails?

\textbf{RQ2:} How does the performance of the supported SMT solvers compare on floating-point queries generated from real-world code?

\textbf{RQ3:} Do NaN-checks on comparisons and type cast safety checks detect possible bugs in real-world code?

\subsection{Experimental Setup}

We use three sets of benchmarks representing real-world floating-point programs:
benchmarks used in the evaluation of \key{}~\cite{abbasiCombiningRuleSMTbased2023},
functions from Stainless' standard math library (\Cref{sec:case_study}), and
floating-point functions extracted from open-source Scala projects on GitHub.


Throughout this section, we consider each annotated source \emph{file} a single benchmark.
Each such benchmark may contain multiple annotated \emph{functions},
each of which may generate multiple verification conditions (\vc{}s).

We run all experiments on Ubuntu~24.02 with an AMD~Ryzen~5900X CPU (12~cores at a 3.7~GHz base clock) and 128~GB of memory.
We use three SMT-solvers: \zthree{}~\cite{deMouraZ32008}~(version 4.13.0), \cvc{}\cite{barbosaCVC52022} (version 1.1.2), and \bitwuzla{}\cite{niemetzBitwuzla2023} (built from latest GitHub commit~\cite{BitwuzlaLatestCommit}).
We run all solvers using a single thread.

For the \key{} benchmarks, we disable Stainless's usability-oriented \code{NaN} checks and safety checks on type casts, since \key{} does not include these features.
For the FdLibm-based benchmarks, we also disable these checks to avoid degraded solver performance due to the larger solver context caused by these checks.
We keep these checks enabled for the empirical GitHub benchmarks.

\subsubsection{\key{} Benchmarks.}

Because Stainless primarily targets functional programs, we adapt the benchmarks to a more functional style.
Specifically, we make classes immutable, rewrite mutating methods to return new instances, and replace small fixed-size arrays by tuples.
We also merge tightly coupled Java source files into single Scala files, resulting in some Scala benchmarks corresponding to multiple classes or files in the original Java suite.

Unlike Stainless, the \key{} verifier supports multiple contracts per method.
We translated such \key{} benchmarks by
(1) adding preconditions common to all contracts as a precondition in Stainless, and
(2) encoding each \key{} contract as an implication in Stainless, combining them into a single conjunctive postcondition while omitting any duplicated preconditions.
If a method in a \key{} benchmark has multiple contracts, one of which is invalid, we avoid using the invalid contract to prove other contracts
by generating two versions of the method: one containing only valid contracts and one including the invalid contract.
No method in the \key{} benchmarks contains more than one invalid contract.

\subsubsection{FdLibm Benchmarks.}

The case study presented in \Cref{sec:case_study} constitutes the second set of benchmarks.
Each function in the math library is converted into one benchmark that requires proving all lemmas about the function,
under the assumption that all lemmas about other functions in the library are correct.
Two exceptions apply: the sine and cosine functions are combined into a single benchmark since they share most of their code,
and the remainder-modulo-$\pi \slash 2$ (RemPio2) operation---used to implement sine, cosine, and tangent functions---is treated as a standalone benchmark.

\subsubsection{Empirical Benchmarks.}

\label{sec:emp_eval}

\begin{figure*}[t]
    \centering
    \includegraphics[width=0.95\textwidth]{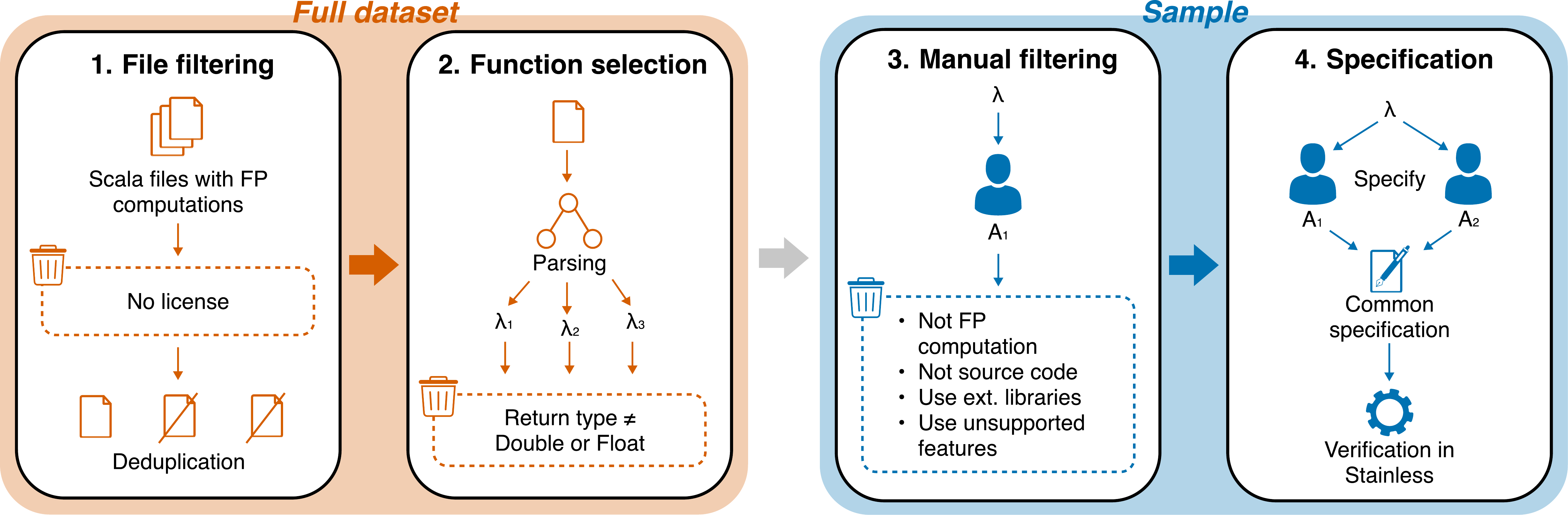}
    \caption{Overview of the methodology to extract benchmarks from GitHub
    }
    \label{fig:benchmarks}
\end{figure*}

Previous work\cite{gilot2025largescalestudyfloatingpointusage} shows that common floating-point verification benchmarks do not accurately represent real-world programs.
In their study, the authors extract a dataset of GitHub source files containing floating-point computations in statically typed
languages. We build on this dataset and use the Scala subset to derive empirical benchmarks for Stainless.
\Cref{fig:benchmarks} outlines our overall methodology for extracting and specifying these benchmarks.

We select all Scala files with open-source licenses, remove duplicates to mitigate large-scale code duplication~\cite{lopesDejaVu2017}, and extract all functions returning floating-point values.
From these, we uniformly sample 500 functions and manually exclude:
\begin{itemize}
    \item 178 functions that return a floating-point value but do not perform any floating-point computation (e.g., simple getters). 
    \item 54 functions that do not belong to the source code of the repository (e.g., to a test suite). 
    \item 97 functions that directly or indirectly call functions from external libraries (e.g., Breeze, Apache Spark). 
    \item 70 functions that use language features unsupported by Stainless, such as macros, concurrency, or bounded polymorphism.
    When possible, we adapt such functions by removing unsupported features without altering their semantics. 
\end{itemize}
After filtering, we obtain a set of 103 functions from 41 distinct repositories.
Since functions may depend on one another, we group functions from the same
source file and their dependencies into single verification benchmarks.

The first two authors independently write specifications for these functions,
using in-code documentation, usage context, references to published algorithms,
and semantic cues (e.g., a variable named \code{count} is assumed to be non-negative).
We systematically examine whether the code can return special values, such as NaN, for valid inputs.
The annotators then compare their specifications and resolve disagreements through discussion. 
When consensus cannot be reached, both specifications are retained and verified independently. 
If Stainless produces a counterexample when verifying one of the benchmarks, we manually correct the implementation and reattempt verification. 
This process yields two distinct benchmarks, one representing the original implementation and one incorporating the correction.
\TightenPar{}

The resulting empirical benchmark suite consists of 79 benchmarks from 52 unique source files;
9 source files had unresolved specification disagreements, and 18 benchmarks were derived from fixing counterexamples.

\subsection{RQ1: Verification Results on Full Benchmarks}
We evaluate the floating-point support in Stainless by running end-to-end experiments on all benchmark sets.
We configure Stainless to use \cvc{}, \bitwuzla{}, and \zthree{} in a portfolio, with a 300-second timeout per \vc{} for the \key{} and the empirical benchmarks.
We do not use any timeout for the more numerically intensive FdLibm benchmarks, as they are designed to establish the correctness of specific lemmas about the library functions, which requires all \vc{}s to terminate.
We do not empirically compare with \key{} or Frama-C (\eg{} measuring performance), as these tools target different programming paradigms. Such a comparison would require extensive modification of the benchmarks, as our translation of the KeY benchmarks demonstrates, making the results difficult to interpret.

Three benchmarks are excluded from the evaluation to prevent execution issues.
Two of them are empirical benchmark: one causes stack-overflows in Stainless' termination inference mechanism,
while the other one causes Stainless to freeze, requiring manual interruption.
We also exclude the FdLibm RemPio2 benchmark, which triggers out-of-memory errors on the evaluation machine.
The benchmark can, however, be successfully verified on the laptop used for the FdLibm case study, where it does not use an unusual amount of memory.  We will investigate these issues further.

\begin{table}
	\fontsize{8}{10}\selectfont   
	\caption{Summary statistics for five runs of the end-to-end experiments.
	The min/median/max columns summarise the end-to-end execution times.
	The column \#VCs shows the number of verification conditions of each benchmark with
	valid(/invalid/unknown/timeout) results in the run with the median execution time. The columns \cvc{},
	\bitwuzla{}, and \zthree{} show the number of VCs shown to be valid(/invalid)
	by each of the solvers in the same median run.
	The column \#fns shows the number of calls to transcendental math library functions, when not zero.}
	\setlength{\tabcolsep}{2pt}
	\begin{subtable}{\textwidth}
		\centering
		\caption{\key{} benchmarks (300s timeout)}
		\label{tab:end2end-key}
		\begin{adjustbox}{width=1\textwidth}
		\begin{tabular}{R{3.8cm}@{\hspace{0.5em}}|R{1cm}R{1cm}R{1cm}R{1.5cm}@{\hspace{0.5em}}|C{0.6cm}C{1.2cm}C{0.5cm}|C{0.7cm}}
			\toprule
			benchmark & min & median & max & \#vcs & \cvc{} & \bitwuzla{} & \zthree{} & \#fns \\
			\midrule
			CartesianPolar & 12.92 & 13.05 & 13.15 & 3 & 3 & 0 & 0 & 3 \\
			Complex & 94.74 & 95.62 & 96.06 & 9/1/0/0 & 9/1 & 0 & 0 & 3 \\
			FPLoop & 4.48 & 4.55 & 4.67 & 30/3/0/0 & 13 & 12/3 & 5 \\
			Matrix2 & 2.50 & 2.53 & 2.61 & 1 & 1 & 0 & 0 \\
			Matrix3 & 129.44 & 130.34 & 130.65 & 1/1/0/0 & 1/1 & 0 & 0 \\
			PineBench & 234.66 & 245.32 & 247.01 & 105/14/0/0 & 58/2 & 46/12 & 1 \\
			Rectangle & 10.53 & 10.55 & 10.72 & 1/1/0/0 & 0/1 & 0 & 1 \\
			Rotation & 21.05 & 21.17 & 21.30 & 6 & 4 & 2 & 0 \\
			\bottomrule
		\end{tabular}
		\end{adjustbox}
	\end{subtable}
	\begin{subtable}{\textwidth}
		\centering
		\caption{FdLibm benchmarks (no timeout)}
		\label{tab:end2end-fdlibm}
		\begin{adjustbox}{width=1\textwidth}
		\begin{tabular}{R{3.8cm}@{\hspace{0.5em}}|R{1cm}R{1cm}R{1cm}R{1.5cm}@{\hspace{0.5em}}|C{0.6cm}C{1.2cm}C{0.5cm}|C{0.7cm}}
			\toprule
			benchmark & min & median & max & \#VCs & \cvc{} & \bitwuzla{} & \zthree{} & \#fns \\
			\midrule
			Acos & 33.72 & 35.60 & 42.28 & 11 & 1 & 6 & 4 \\
			Asin & 105.46 & 171.82 & 859.33 & 8 & 1 & 6 & 1 \\
			Atan & 62.43 & 64.44 & 70.88 & 10 & 0 & 8 & 2 \\
			Atan2 & 8.40 & 8.47 & 8.56 & 28 & 0 & 7 & 21 & 2 \\
			Cbrt & 667.36 & 913.66 & 1178.50 & 11 & 1 & 7 & 3 \\
			Cosh & 40.07 & 41.85 & 45.40 & 9 & 1 & 7 & 1 & 4 \\
			Exp & 49.85 & 52.04 & 152.58 & 16 & 0 & 11 & 5 \\
			Expm1 & 3224.93 & 3969.72 & 4189.57 & 24 & 1 & 19 & 4 \\
			Hypot & 16.88 & 17.49 & 18.24 & 7 & 0 & 6 & 1 \\
			Log & 140.12 & 206.87 & 741.37 & 16 & 1 & 14 & 1 \\
			Log10 & 4.10 & 4.21 & 4.22 & 6 & 0 & 4 & 2 & 1 \\
			Log1p & 22.78 & 120.52 & 249.31 & 16 & 1 & 13 & 2 \\
			Pow & 9.40 & 9.57 & 9.63 & 24 & 1 & 21 & 2 \\
			SinCos & 194.80 & 220.15 & 511.64 & 42 & 12 & 3 & 27 \\
			Sinh & 111.82 & 119.59 & 125.88 & 9 & 1 & 8 & 0 & 3 \\
			Tan & 17.80 & 17.93 & 18.09 & 13 & 3 & 4 & 6 \\
			Tanh & 4.35 & 4.42 & 4.59 & 6 & 0 & 5 & 1 & 2 \\
			\bottomrule
		\end{tabular}
		\end{adjustbox}
	\end{subtable}
	\begin{subtable}{\textwidth}
		\centering
		\caption{Subset of empirical benchmarks uniformly selected by median time (300s timeout)}
		\label{tab:end2end-empirical}
		\begin{adjustbox}{width=1\textwidth}
		\begin{tabular}{R{3.8cm}@{\hspace{0.5em}}|R{1cm}R{1cm}R{1cm}R{1.5cm}@{\hspace{0.5em}}|C{0.6cm}C{1.2cm}C{0.5cm}|C{0.7cm}}
			\toprule
			benchmark & min & median & max & \#VCs & \cvc{} & \bitwuzla{} & \zthree{} & \#fns \\
			\midrule
			HTMLComp.Position\_OK & 3.30 & 3.35 & 3.43 & 2 & 1 & 1 & 0 \\
			StreamingIO\_CE & 3.58 & 3.60 & 3.65 & 2/1/0/0 & 1/1 & 1 & 0 \\
			BoundingBox\_a1\_OK & 3.85 & 3.89 & 4.00 & 4 & 2 & 0 & 2 \\
			AreaUnderCurve\_OK & 4.16 & 4.21 & 4.28 & 11 & 10 & 0 & 1 \\
			DataHelper\_OK & 4.69 & 4.79 & 4.88 & 21 & 4 & 0 & 17 \\
			Player\_a0\_OK & 6.26 & 6.38 & 6.54 & 21 & 16 & 0 & 5 \\
			Border\_OK & 7.74 & 7.83 & 7.87 & 77 & 14 & 0 & 63 \\
			Mediator\_a1\_OK & 9.41 & 9.50 & 9.71 & 56 & 10 & 0 & 46 \\
			Variance\_CE & 15.25 & 15.44 & 15.61 & 37/2/0/0 & 18/2 & 4 & 15 \\
			Vector2\_a0\_CE & 33.42 & 33.55 & 33.75 & 7/2/0/0 & 3/2 & 0 & 4 \\
		    Common\_TO & 727.42 & 748.05 & 916.92 & 46/0/1/2 & 15 & 5 & 26 & 1 \\
			\bottomrule
		\end{tabular}
		\end{adjustbox}
	\end{subtable}
	
	\label{tab:end2end-summary}
\end{table}

\Cref{tab:end2end-summary} summarises the evaluation results for the \key{} and FdLibm benchmarks, and a subset of the empirical benchmarks.
Comprehensive statistics for all empirical benchmarks are provided in
\ifarxiv
\Cref{tab:end2end-empirical-full} in the appendix.
\fi
For each benchmark, we report the minimum, median, and maximum Stainless end-to-end wall-clock execution times over five independent runs.
We also include the total number of valid and invalid \vc{}s, along with the numbers of \vc{}s reported as unknown or resulting in timeouts.
Additionally, for the median run, we report the number of valid and invalid \vc{}s solved by each solver.

For the \key{} and FdLibm benchmark sets, all \vc{}s are successfully shown to be either valid or invalid.
For the empirical benchmarks, Stainless reports at least one unknown result or timeout in 6 out of 79 empirical cases: 3 benchmarks experience at least one unknown result, and 4 encounter at least one timeout.
All \vc{}s are successfully shown to be either valid or invalid for 73 out of 79 empirical benchmarks.
\TightenPar{}

Manual inspection reveals that for 8 empirical benchmarks detected as invalid by Stainless, the reported counterexamples are spurious.%
\footnote{For one particular run, the SMT-solvers are not guaranteed to produce the same counterexamples in different runs.}
These benchmarks contain transcendental functions, which Stainless treats as opaque. That is, only their contract is available to the rest of the codebase and its implementation is hidden. The function call is translated to the SMT solver as an application of an uninterpreted function constrained solely by the contract. If the contract is not strong enough to discharge a verification condition, the solver can construct a model that satisfies the contract while violating the intended semantics of the function, leading to a spurious counterexample.
Among the other 11 empirical benchmarks using transcendental functions, one yields a timeout, 5 are verified, and 5 yield non-spurious counterexamples.
Overall, the axiomatic approach introduced by \key{} is insufficient for 8 out of 19 benchmarks using transcendental functions, \ie{} a significant fraction.

\begin{rqanswer}
\textbf{Answer to RQ1:} Our results indicate that Stainless is usually able to successfully verify or produce counterexamples for self-contained real-world floating-point programs (82 \% of empirical cases).
In some cases, we observe long verification times with a high variance, especially for the more numerically intensive FdLibm benchmarks.
Transcendental functions still pose a challenge:
the lemmas introduced in the case study over-approximate their behaviour, leading to spurious counterexamples in several benchmarks (8 out of 19).
\end{rqanswer}

\subsection{RQ2: Comparison of Solvers on Verification Conditions}\label{sec:eval-rq2}


\begin{table}[t] 
	\caption{Number of verification conditions solved by each solver}\label{tab:solved-vcs}
	\fontsize{8}{10}\selectfont   
	\setlength{\tabcolsep}{4pt}
	\centering
	\begin{tabular}{lr@{\hspace{1em}}|@{\hspace{1em}}llll}
		\toprule
		          & \#VCs & \cvc{}           & \zthree{}            & \bitwuzla{}     & Any solver    \\
		\midrule
		KeY       & 176   & 176 (100\%)    & 150 (85.2\%)  & 118 (67.0\%) & 176 (100\%)   \\
		FdLibm    & 256   & 227 (88.7\%)   & 200 (78.1\%)  & 217 (84.8\%) & 254 (99.2\%)  \\
		Empirical & 1600  & 1595 (99.7\%)  & 1528 (95.5\%) & 489 (30.6\%) & 1596 (99.8\%) \\
		\bottomrule
	\end{tabular}
\end{table}

To evaluate and compare the performance of the SMT solvers supported by Stainless, we configure Stainless to generate one SMT-LIB file per \vc{} in each benchmark.
We run each solver on each generated SMT-LIB file five times and measure the total execution time.
For \vc{}s generated from the \key{} and FdLibm benchmarks, we use a timeout of 300 seconds.
Given the large number of empirical benchmarks, we use a reduced timeout of 120 seconds for this set.

\Cref{tab:solved-vcs} reports the number of \vc{}s solved by \cvc{}, \zthree{} and \bitwuzla{} across all benchmark categories.
The table also lists the total number of VCs solved by at least one of the three solvers.
Compared to the end-to-end evaluation, we observe fewer unsolved \vc{}s, since unknown results can originate from Stainless itself;
in such case we do not observe any unknown results from an SMT-solver.
The results indicate that the portfolio approach is effective, with almost all \vc{}s being solved by at least one solver.

\begin{figure}[t]
	\centering
	\begin{subfigure}[t]{\textwidth}
		\includegraphics[width=\textwidth]{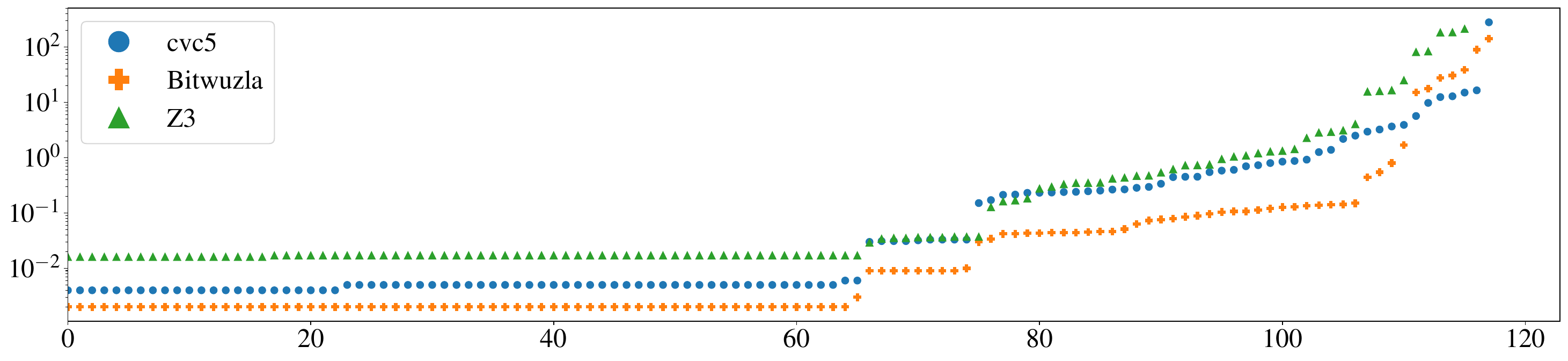}
		\caption{\key{} benchmarks (300s timeout)}
		\label{fig:cactus-key-solved}
	\end{subfigure}
	\begin{subfigure}[t]{\textwidth}
		\includegraphics[width=\textwidth]{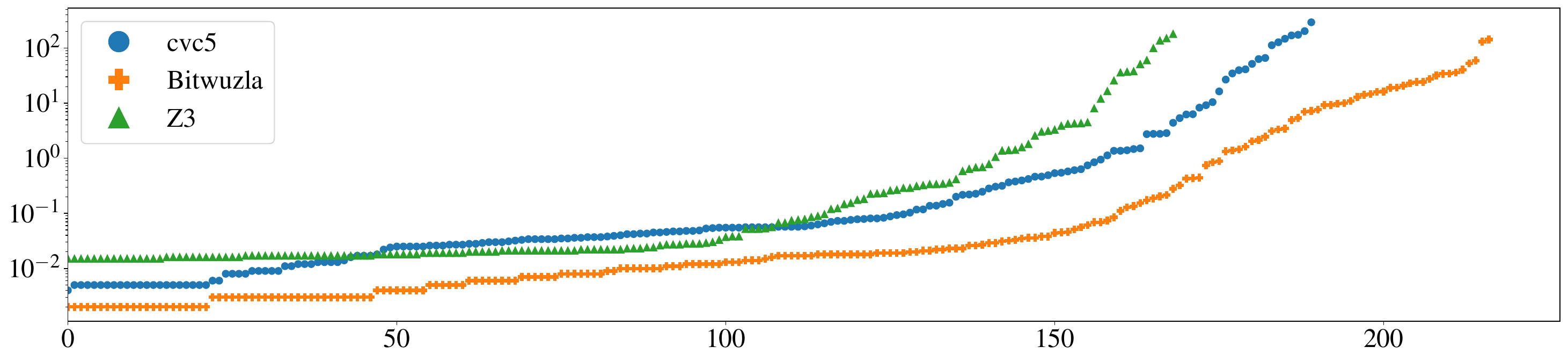}
		\caption{FdLibm benchmarks (300s timeout)}
		\label{fig:cactus-fdlibm-solved}
	\end{subfigure}
	\begin{subfigure}[t]{\textwidth}
		\includegraphics[width=\textwidth]{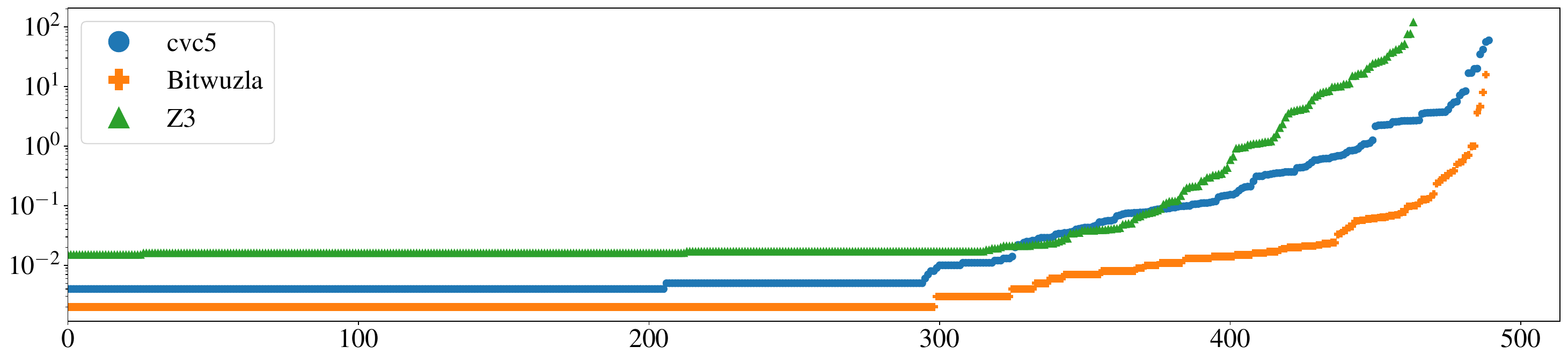}
		\caption{empirical benchmarks (120s timeout)}
		\label{fig:cactus-empirical-solved}
	\end{subfigure}
	\caption{Solver runtime for verification conditions supported by all solvers}
	\label{fig:cactus-solved}
\end{figure}

Most unsolved \vc{}s for \bitwuzla{} correspond to errors rather than performance limitations:
it yields errors for 58 \key{} benchmarks, 37 FdLibm benchmarks, and 1107 empirical benchmarks while \cvc{} and \zthree{} do not encounter any errors.
We believe all of these errors are due to a lack of support for algebraic data types and the theory of integers.
We therefore restrict the comparison between the solvers to the subset of \vc{}s where \bitwuzla{} does not encounter any errors.
\Cref{fig:cactus-solved} visualises the performance of the three solvers on this subset, and
\ifarxiv
\Cref{fig:cactus-all} in the appendix
\fi
contains corresponding plots for all \vc{}s.

\begin{rqanswer}
\textbf{Answer to RQ2:}
We get a clear ranking of the solvers for the cases where \bitwuzla{} does not encounter any errors:
\bitwuzla{} performs better than \cvc{}, especially for the numerically intensive FdLibm benchmarks, and \cvc{} performs better than \zthree{}.
Since \bitwuzla{} does not support algebraic data types or the theory of integers, \cvc{} still solves more \vc{}s overall.
\end{rqanswer}

\subsection{RQ3: Impact of NaN-Checks and Type Cast Safety Checks}\label{sec:eval-rq3}

Among the 103 functions in our empirical benchmark set, 9 contain floating-point comparisons.
In all these cases, proper handling of \code{NaN} values is missing.
In two functions, this omission causes the implementation to return a valid output even when the input is \code{NaN}, effectively silently ignoring \code{NaN} inputs.
In one case the code comments explicitly assume that a false comparison implies that the input is valid and within a certain range, a property that does not hold when the input is \code{NaN}; we show this function in
\ifarxiv
\Cref{lst:comparison-nan} in the appendix.
\fi

In most of the functions, function arguments are compared against constant thresholds.
In such cases, the assumption that inputs are non-\code{NaN} is sufficient to guarantee the absence of erroneous comparisons.
When a function is lacking this assumption, Stainless successfully detects and reports potential \code{NaN} comparisons.

Floating-point to integer casts occur in two functions within the benchmark set.
Both are unsafe, and are reported by Stainless with concrete counterexamples that trigger either \code{NaN}-to-integer conversions or out-of-bounds integer casts.
\begin{rqanswer}
\textbf{Answer to RQ3:} Automated \code{NaN}-checks and type-cast safety checks effectively identify floating-point-related bugs in user code.
In our empirical benchmark set, Stainless successfully detects all unsafe type casts and potential \code{NaN} comparisons, generating concrete counterexamples for each such case.\TightenPar{}
\end{rqanswer}


\section{Related Work}
\label{sec:related_work}

To the best of our knowledge, we are the first to introduce floating-point
reasoning into a deductive verifier for Scala.
%
Stainless' floating-point support via verification conditions discharged by SMT
solvers using the floating-point SMT-LIB theory~\cite{rummer2010smt} follows
the support available in KeY for Java, Why3~\cite{filliatreWhy3WhereProgramsMeetProvers2013} for
C~\cite{cuoqFramaCSourceCode2012} and Ada~\cite{fumexAutomatingVerification2017},
and the Boogie intermediate language~\cite{leinoBoogie2008}.
Compared to these tools, Stainless provides floating-point support for an input
language with a different set of features that include polymorphism, recursion,
higher-order functions and class invariants. It also adds automated \code{NaN}
checks, reducing the annotation burden compared to existing verifiers.
KeY introduced an axiomatic treatment of some transcendental functions from the
standard math library~\cite{abbasiCombiningRuleSMTbased2023}. In comparison,
Stainless covers the entire standard math library and, for most functions, is
able to verify the axioms against the actual implementation itself.
For other JVM-based languages, deductive verifiers provide only limited support:
VeriFast~\cite{jacobsVeriFast2011} treats floating-point operations as reals,
while OpenJML~\cite{cokOpenJMLJMLJava2011} and LiquidJava~\cite{gamboaUsabilityOrientedDesign2023} can parse floating-point programs
but do not reason about them.
\TightenPar

Abstract interpretation-based
techniques~\cite{chenSoundFloatingPoint2008,jeannetApronLibraryNumerical2009}
can in principle automatically prove the absence of special values in
floating-point code. However, applying these analyses effectively to real-world
programs typically requires significant user tuning, such as selecting abstract
domains or widening thresholds~\cite{blanchetStaticAnalyzerLarge2003}.
Ariadne~\cite{barrAutomaticDetectionFloatingPoint2013} uses a combination of
symbolic execution, real-valued SMT solving and testing to find inputs that
trigger floating-point exceptions, including overflow and invalid operations.
Our work subsumes this line of analysis: Stainless not only generates
counterexamples through SMT solving, but also proves the absence of such
exceptions.
\TightenPar{}

Besides showing the absence of special values, recent research has developed
static analyses to bound floating-point rounding errors
\cite{moscatoAutomaticEstimationVerified2017,magronCertifiedRoundoffError2017,
solovyevRigorousEstimationFloatingPoint2018,darulovaDaisyFrameworkAnalysis2018,goubaultStaticAnalysisFinite2011}.
These analyses, however, currently target small arithmetic kernels expressed in
domain-specific languages, and lack support for features of mainstream
programming languages such as user-defined data types, objects, and recursion.

Stainless verifies the axioms for the standard math library functions against
the implementations using its own floating-point support. Bagnara et
al.~\cite{bagnaraPracticalApproachVerification2021} verify properties of
functions in the \code{math.h} C library, such as piecewise monotonicity and the
absence of \code{NaN} results under valid inputs using an alternative approach
that combines symbolic execution, abstract interpretation, and exhaustive
testing.
A complementary line of research verifies the error bounds for
floating-point implementations of transcendental
functions in interactive
theorem provers (\eg{} HOL, Rocq)~\cite{harrisonFormalVerification2000,harrisonFloatingPointVerification2000,
harrisonFormalVerification2003,akbarpourVerifyingSynthesizedImplementation2009,boldoCombiningCoqGappa2009,
melquiondFloatingpointArithmeticCoq2012}; such proofs require an extensive mathematical
apparatus and significant user expertise and interaction. Harrison reports that
the manual effort required for each such proof can vary from weeks to months
\cite{harrisonFloatingPointVerification2000}.

\section{Conclusion}
\label{sec:conclusion}

Supporting floating-point reasoning in a deductive verifier for programs that use advanced language features such as classes, polymorphism, and higher-order functions requires specific restrictions, for example by limiting instantiations of polymorphic equality or by weakening equality through uninterpreted functions.
With our extension of Stainless, we reproduced verification results of KeY and extended them by verifying a large subset of the Scala standard math library, which KeY previously axiomatised or did not support, and showed that Stainless supports verification of user code sampled from GitHub.


Future work includes detecting and eliminating spurious counterexamples arising from transcendental functions.
When spurious examples are detected (\eg{} by evaluating the program), one possible strategy is to inline the full function implementation and reattempt verification.
Current SMT solvers, however, are generally not able to reason efficiently about the complete implementation of transcendental functions, making this approach often ineffective in practice.
Strengthening the contracts with additional properties can mitigate the problem, although it does not eliminate the underlying issue completely.
Future work also includes investigating usability trade-offs, for example by adding further automated checks and designing more readable specification mechanisms.

\begin{credits}
\subsubsection{\ackname}
Co-funded by the European Union (ERC, HORNET, 101163629). Views and opinions expressed are however those of the author(s) only and do not necessarily reflect those of the European Union or the European Research Council. Neither the European Union nor the granting authority can be held responsible for them.
\TightenPar{}

\subsubsection{\discintname}
The authors have no competing interests to declare that are relevant to the content of this article.
\end{credits}

\section{Data Availability Statement}
Our floating-point support for Stainless is available in the main branch of the Stainless repository: \url{https://github.com/epfl-lara/stainless}.
We also provide a Zenodo artifact (\url{https://doi.org/10.5281/zenodo.17486575}) containing the implementation, all benchmarks used in this paper, the collected evaluation results, and scripts to reproduce the experiments.

\bibliographystyle{splncs04}
\bibliography{bibliography/float-verification_minimal}

\ifarxiv

\newpage
\appendix
\section*{Appendix}

\begin{table}[ht]
\caption{Comparison of transcendental function axioms in KeY and verified properties in Stainless}
\label{tab:key_vs_stainless_axioms}
\centering
    \setlength{\tabcolsep}{3pt}
    \renewcommand{\arraystretch}{1.2}
    \begin{adjustbox}{width=1\textwidth}
    \begin{tabular}{lll}
    \textbf{Function} & \textbf{KeY (Axioms)} & \textbf{Stainless (Verified)} \\
    \toprule
                    cos      & NaN, range                     & NaN, range \\
\rowcolor{black!5}  sin      & NaN, f($\pm 0$), range         & NaN, f($\pm 0$), range \\
                    tan      & f(NaN), f($\pm 0$), range      & f(NaN), f($\pm 0$), range \\
\rowcolor{black!5}  asin     & f(NaN), f(0), range            & NaN, f($\pm 0$), f($\pm 1$), range \\
                    acos     & f(NaN), f(0), range            & NaN, f($\pm 0$), f($\pm 1$)\\
\rowcolor{black!5}  atan     & f(NaN), f(0), range            & NaN, f($\pm 0$), f($\pm \infty$), sign, range\\
                    atan2    & f(NaN, \_), f(\_, NaN), range  & NaN, all pairs of special values, sign, range\\
\rowcolor{black!5}  hypot    & Not supported                  & NaN, all pairs of special values, range \\
                    cbrt     & Not supported                  & NaN, f($\pm 0$), f($\pm 1$), f($\pm \infty$), sign, $f(x) \lessgtr x$ \\
\rowcolor{black!5}  pow      & All pairs of special values    & All pairs of special values \\
                    exp      & f(NaN), f($\pm \infty$)        & NaN, f($\pm 0$), f($\pm \infty$), range \\
\rowcolor{black!5}  expm1    & Not supported                  & NaN, f($\pm 0$), f($\pm \infty$), range \\
                    log      & Not supported                  & NaN, f($\pm 0$), f($1$), f($\pm \infty$), range, $f(x) \leq x - 1$ \\
\rowcolor{black!5}  log1p    & Not supported                  & NaN, f($\pm 0$), f($- 1$), f($\pm \infty$), range, $f(x) \leq x$ \\
                    log10    & Not supported                  & NaN, f($\pm 0$), f($1$), f($\pm \infty$), range \\
\rowcolor{black!5}  sinh     & Not supported                  & NaN, f($\pm 0$), f($\pm \infty$), sign\\
                    cosh     & Not supported                  & NaN, f($\pm 0$), f($\pm \infty$), range\\
\rowcolor{black!5}  tanh     & Not supported                  & NaN, f($\pm 0$), f($\pm \infty$), sign, range\\
    \toprule
\multicolumn{3}{l}{\textbf{Legend:}}\\
\multicolumn{3}{l}{\textit{NaN}: exact characterisation of when the output of the function is \texttt{NaN}}\\
\multicolumn{3}{l}{$f(\square)$: expected result for input $\square$}\\
\multicolumn{3}{l}{\textit{all pairs of special values}: expected result for all combinations of \texttt{NaN}, $\pm 0$, and $\pm \infty$}\\
\multicolumn{3}{l}{\textit{sign}: characterisation of the sign of the result}\\
\multicolumn{3}{l}{\textit{range}: lower and/or upper bounds on the result}\\
\multicolumn{3}{l}{\textit{$f(x) \lessgtr x$}: characterisation of when the function is above or below its input}\\
    \end{tabular}
    \end{adjustbox}
\end{table}


\newpage
	\setlength{\tabcolsep}{2pt}
	\fontsize{8}{10}\selectfont   
	\begin{longtable}{rrrrrrrrr}
		\caption{Summary statistics for five runs of the end-to-end experiments on the empirical benchmarks.  
		The min/median/max columns summarise the end-to-end execution times.
		 The column \#vcs shows the total number of verification conditions shown to be valid in the median run, or, 
		 if some were not shown to be valid, the number of valid/invalid/unknown/timeout results.  
		 The columns \cvc{}, \bitwuzla{}, and \zthree{} shows the number of verification conditions shown to be valid or the number shown to be valid/invalid by each solver in the median run.  
		 The column \#fns shows the number of calls to transcendental math library functions, when not zero.} \\
		\toprule
		benchmark & min & median & max & \#vcs & \cvc{} & \bitwuzla{} & \zthree{} & \#fns \\
		\midrule
		AbsoluteError\_OK & 3.35 & 3.37 & 3.41 & 2 & 0 & 2 & 0 \\
		AreaUnderCurve\_CE & 4.18 & 4.22 & 4.38 & 9/1/0/0 & 7/1 & 0 & 2 \\
		AreaUnderCurve\_OK & 4.16 & 4.21 & 4.28 & 11 & 10 & 0 & 1 \\
		Border\_OK & 7.74 & 7.83 & 7.87 & 77 & 14 & 0 & 63 \\
		BoundingBox\_a0\_OK & 4.01 & 4.07 & 4.11 & 6 & 4 & 0 & 2 \\
		BoundingBox\_a1\_OK & 3.85 & 3.89 & 4.00 & 4 & 2 & 0 & 2 \\
		Cast\_OK & 3.57 & 3.59 & 3.63 & 4 & 0 & 4 & 0 \\
		Common\_TO & 727.42 & 748.05 & 916.92 & 46/0/1/2 & 15 & 5 & 26 & 1 \\
		DataHelper\_OK & 4.69 & 4.79 & 4.88 & 21 & 4 & 0 & 17 \\
		DataSetCost\_false\_CE & 32.33 & 32.47 & 32.71 & 12/2/0/0 & 2/2 & 0 & 10 & 1 \\
		Exp.Mutagen\_false\_CE & 16.97 & 17.13 & 17.30 & 11/2/0/0 & 2/2 & 0 & 9 & 1 \\
		FlinkStatistic\_OK & 4.24 & 4.29 & 4.34 & 10 & 5 & 0 & 5 \\
		Float16\_a0\_OK & 4.27 & 4.34 & 4.38 & 17 & 7 & 0 & 10 \\
		Float16\_a1\_OK & 4.50 & 4.58 & 4.62 & 21 & 16 & 2 & 3 \\
		FollowTargetRule\_CE & 100.47 & 100.71 & 100.76 & 15/9/0/0 & 10/9 & 1 & 4 & 4 \\
		FollowTargetRule\_OK & 104.89 & 104.91 & 105.39 & 32/3/0/0 & 20/3 & 2 & 10 & 4\\
		GCode\_a0\_CE & 3.86 & 3.88 & 3.95 & 2/1/0/0 & 0/1 & 1 & 1 \\
		GCode\_a0\_OK & 3.92 & 3.94 & 4.00 & 4 & 1 & 1 & 2 \\
		GCode\_a1\_OK & 3.80 & 3.90 & 4.00 & 3 & 1 & 0 & 2 \\
		HTMLComp.Position\_OK & 3.30 & 3.35 & 3.43 & 2 & 1 & 1 & 0 \\
		HyperLogLogPlusPlus\_CE & 46.40 & 46.77 & 46.88 & 4/4/0/0 & 2/4 & 0 & 2 & 2 \\
		Intro\_OK & 3.40 & 3.51 & 3.53 & 2 & 0 & 2 & 0 \\
		IsotonicRegression\_CE & 5.02 & 5.11 & 5.68 & 15/1/0/0 & 1 & 0/1 & 14 \\
		IsotonicRegression\_OK & 6.43 & 6.70 & 7.53 & 31 & 5 & 11 & 15 \\
		JaccardSimilarity\_OK & 10.48 & 10.63 & 10.78 & 16 & 12 & 0 & 4 \\
		JaccardSimilarity\_UNK & 104.82 & 109.56 & 428.60 & 10/0/2/0 & 6 & 0 & 4 \\
		Kitchen\_OK & 4.70 & 4.80 & 4.98 & 5 & 1 & 0 & 4 \\
		Kmath\_a0\_OK & 8.04 & 8.25 & 8.48 & 12 & 4 & 5 & 3 & 3 \\
		Kmath\_a1\_OK & 8.68 & 8.77 & 8.82 & 12 & 4 & 5 & 3 & 3 \\
		LDAOptimizer\_false\_CE & 6.68 & 6.77 & 6.89 & 3/2/0/0 & 0/2 & 0 & 3 & 1 \\
		List\_OK & 6.95 & 7.03 & 7.21 & 18 & 4 & 0 & 14 \\
		LogLoss\_CE & 6.90 & 6.99 & 7.10 & 24/2/0/0 & 1 & 8/2 & 15 & 6 \\
		LogLoss\_OK & 6.69 & 6.83 & 6.87 & 26 & 2 & 9 & 15 & 6 \\
		MCOpt.DeepStructure\_OK & 3.60 & 3.68 & 3.71 & 4 & 4 & 0 & 0 \\
		Mediator\_a0\_OK & 4.83 & 4.89 & 5.00 & 13 & 4 & 0 & 9 \\
		Mediator\_a0\_UNK & 6.72 & 6.91 & 7.03 & 7/0/3/0 & 1 & 0 & 6 \\
		Mediator\_a1\_OK & 9.41 & 9.50 & 9.71 & 56 & 10 & 0 & 46 \\
		NewtonSquare\_CE & 54.00 & 54.47 & 54.64 & 19/3/0/0 & 6 & 13/3 & 0 \\
		OntologyStatistics\_a0\_OK & 3.51 & 3.56 & 3.68 & 5 & 1 & 4 & 0 \\
		OntologyStatistics\_a1\_OK & 3.73 & 3.84 & 3.91 & 2 & 0 & 2 & 0 \\
		Optimizer\_OK & 3.67 & 3.71 & 3.79 & 7 & 1 & 1 & 5 \\
		PairwiseScorer\_OK & 5.39 & 5.49 & 5.56 & 18 & 6 & 0 & 12 \\
		Player\_a0\_OK & 6.26 & 6.38 & 6.54 & 21 & 16 & 0 & 5 \\
		Player\_a1\_OK & 7.65 & 7.87 & 8.03 & 29 & 26 & 0 & 3 \\
		Polyn.Mutagen\_false\_CE & 14.29 & 14.37 & 14.53 & 9/2/0/0 & 1/2 & 0 & 8 & 2 \\
		PolynomialPerformance\_CE & 4.33 & 4.38 & 4.40 & 12/2/0/0 & 0 & 4/2 & 8 \\
		PolynomialPerformance\_TO & 430.95 & 432.50 & 435.50 & 16/0/0/1 & 1 & 6 & 9 \\
		PredictionMetrics\_CE & 19.13 & 19.14 & 19.29 & 11/2/0/0 & 2/2 & 4 & 5 & 2 \\
		PredictionMetrics\_OK & 15.69 & 15.83 & 16.10 & 17 & 4 & 0 & 13 & 2 \\
		Profile\_CE & 3.99 & 4.11 & 4.19 & 4/3/0/0 & 3/1 & 0/2 & 1 \\
		Profile\_OK & 4.03 & 4.10 & 4.29 & 9 & 4 & 1 & 4 \\
		Rectangle\_OK & 28.62 & 28.70 & 28.81 & 12 & 11 & 0 & 1 \\
		RectangularBorder\_a0\_OK & 11.86 & 12.28 & 12.34 & 30 & 12 & 8 & 10 \\
		RectangularBorder\_a1\_OK & 7.90 & 8.02 & 8.07 & 27 & 15 & 6 & 6 \\
		RedisStore\_OK & 3.53 & 3.59 & 3.65 & 4 & 2 & 2 & 0 \\
		RegularPayGrantCalc.\_OK & 4.39 & 4.43 & 4.51 & 14 & 2 & 9 & 3 \\
		SamplingUtils\_CE & 12.52 & 12.80 & 13.76 & 46/2/0/0 & 11 & 33/2 & 2 & 2 \\
		SamplingUtils\_false\_CE & 13.56 & 14.02 & 14.62 & 50/1/0/0 & 19 & 29/1 & 2 & 2 \\
		Session\_CE & 30.77 & 33.12 & 35.06 & 21/2/0/0 & 6 & 15/2 & 0 \\
		SpeedControl\_OK & 5.56 & 5.61 & 5.76 & 65 & 24 & 41 & 0 \\
		Spherical\_false\_CE & 21.40 & 23.21 & 23.29 & 73/11/0/0 & 21 & 46/11 & 6 & 29 \\
		StatCounter\_OK & 318.74 & 318.82 & 318.93 & 19/0/0/1 & 8 & 0 & 11 \\
		StatCounter\_TO & 318.73 & 318.81 & 319.04 & 17/1/0/1 & 6/1 & 0 & 11 \\
		StormDay\_CE & 3.95 & 3.98 & 4.02 & 2/3/0/0 & 0/2 & 0 & 2/1 \\
		StormDay\_OK & 4.81 & 4.92 & 4.97 & 7 & 3 & 0 & 4 \\
		StreamingHistogram\_CE & 8.68 & 8.80 & 8.95 & 10/1/0/0 & 1/1 & 0 & 9 \\
		StreamingHistogram\_OK & 8.78 & 8.85 & 9.53 & 15 & 5 & 0 & 10 \\
		StreamingIO\_CE & 3.58 & 3.60 & 3.65 & 2/1/0/0 & 1/1 & 1 & 0 \\
		StreamingIO\_OK & 3.38 & 3.47 & 3.54 & 3 & 1 & 2 & 0 \\
		Variance\_CE & 15.25 & 15.44 & 15.61 & 37/2/0/0 & 18/2 & 4 & 15 \\
		Variance\_OK & 13.81 & 13.93 & 14.12 & 43 & 19 & 4 & 20 \\
		Vector2\_a0\_CE & 33.42 & 33.55 & 33.75 & 7/2/0/0 & 3/2 & 0 & 4 \\
		Vector2\_a0\_OK & 33.64 & 33.72 & 34.12 & 16 & 8 & 3 & 5 \\
		Vector2\_a1\_CE & 34.43 & 34.67 & 34.77 & 6/3/0/0 & 2/3 & 1 & 3 \\
		Vector2\_a1\_false\_CE & 6.08 & 6.12 & 6.23 & 18/1/0/0 & 6/1 & 0 & 12 & 1 \\
		Vectors\_false\_CE & 42.87 & 43.45 & 44.02 & 201/3/0/0 & 64/1 & 0 & 137/2 & 2 \\
		WeightedLeastSquares\_OK & 15.87 & 15.90 & 16.03 & 17 & 7 & 0 & 10 \\
		gui\_OK & 3.51 & 3.63 & 3.65 & 3 & 3 & 0 & 0 \\
		package\_OK & 3.73 & 3.79 & 3.80 & 6 & 6 & 0 & 0 \\
		\bottomrule
		\label{tab:end2end-empirical-full}
	\end{longtable}


\begin{figure}[t]
	\centering
	\begin{subfigure}[t]{\textwidth}
		\includegraphics[width=\textwidth]{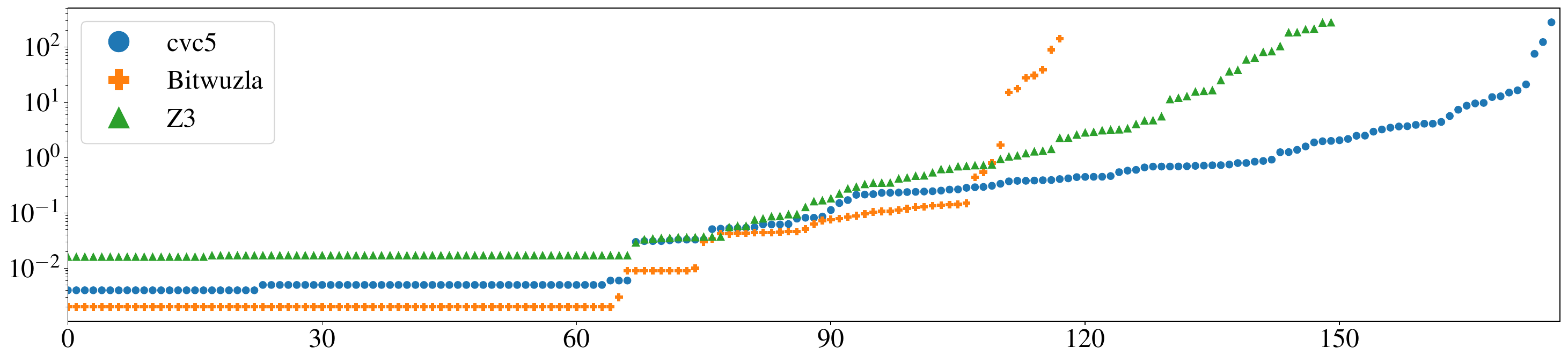}
		\caption{\key{} benchmarks (300s timeout)}
		\label{fig:cactus-key-all}
	\end{subfigure}
	\begin{subfigure}[t]{\textwidth}
		\includegraphics[width=\textwidth]{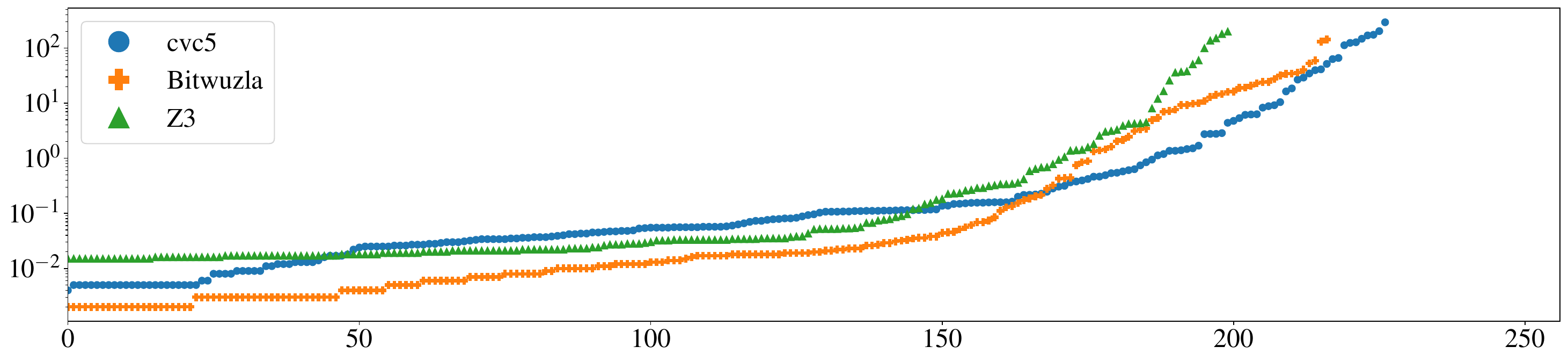}
		\caption{FdLibm benchmarks (300s timeout)}
		\label{fig:cactus-fdlibm-all}
	\end{subfigure}
	\begin{subfigure}[t]{\textwidth}
		\includegraphics[width=\textwidth]{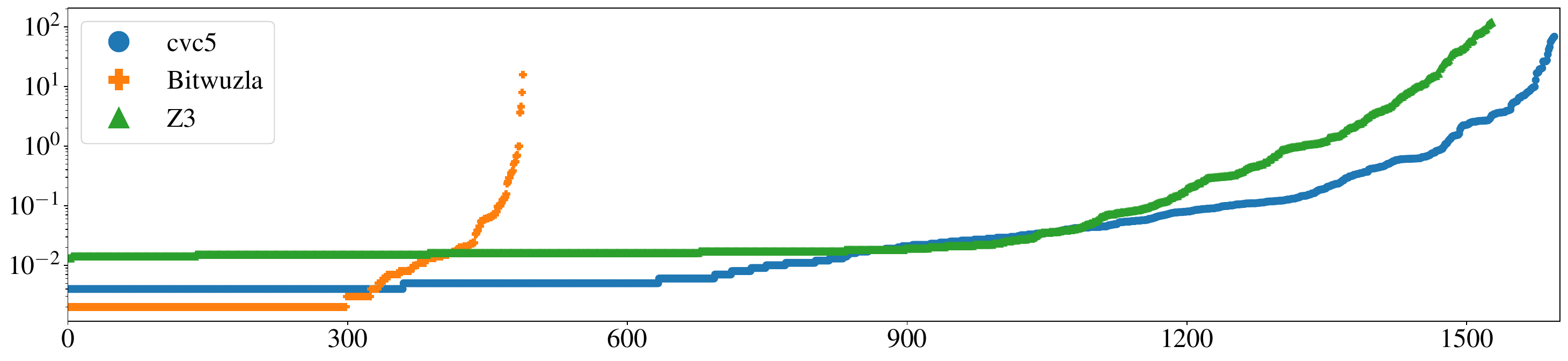}
		\caption{empirical benchmarks (120 second timeout)}
		\label{fig:cactus-empirical-all}
	\end{subfigure}
	\caption{Solver runtime for all verification conditions}
	\label{fig:cactus-all}
\end{figure}

\begin{figure}[t]
\begin{lstlisting}
def getPortionOutsideBorder(position: Double, minReachablePosition: Double, maxReachablePosition: Double): Double = {
	position match {
		case p if p < minReachablePosition => minReachablePosition - p
		case p if p > maxReachablePosition => p - maxReachablePosition
		case _ => 0 // the entity is inside the map edge => radius portion outside the map is 0
	}
}
\end{lstlisting}
\caption{Example of user code with floating-point comparisons where an in-code comment incorrectly assumes that a false comparison implies a valid (non-\code{NaN}) input} \label{lst:comparison-nan}
\end{figure}

\else

\fi



\end{document}